\documentclass[aip,apl,amsmath,amssymb,reprint,floatfix]{revtex4-1}
\usepackage{graphicx}
\usepackage{amsthm}
\usepackage{amssymb,color}
\newcommand{\bl}[1]{{\color{black} #1}}

\newcommand{\bu}[1]{{\color{black} #1}}
\newtheoremstyle{query}%
{}{}
{\color{red}}
{}
{\sffamily\bfseries}{:}{12pt}
{}
\theoremstyle{query}

\begin{document}
\title{Optimized fringe removal algorithm for absorption images}

\author{Linxiao Niu}
\author{Xinxin Guo}
\author{Yuan Zhan}
\author{Xuzong Chen}
\affiliation{School of Electronics Engineering and Computer Science, Peking University, Beijing 100871, China}
\author{W. M. Liu}
\affiliation{Institute of Physics, Chinese Academy of Sciences, Beijing 100080, China}
\author{Xiaoji Zhou}
\email{ xjzhou@pku.edu.cn}
\affiliation{School of Electronics Engineering and Computer Science, Peking University, Beijing 100871, China}

\begin{abstract}

Optical absorption imaging is a basic detection technique for obtaining information from matter waves, in which the absorption signal can be obtained by comparing the recorded detection light field with the light field in the presence of absorption, thereby giving the spatial distribution of the atoms.
The noise in detection arises mainly from differences between the two recorded light field distributions, which is difficult to avoid in  experiments.
In this work, we present an optimized fringe removal algorithm, developing a method to generate an ideal reference light field, avoiding the noise generated  by the light field difference, and suppressing the noise signal to the theoretical limit. Using  principal component analysis, we explore the optimal calculation area and how to remove noise information from the basis to allow optimal performance and speed. As an example, we consider scattering atomic peaks with a small number of atoms in a triangular lattice. Compared with the conventional processing method, our algorithm can reduce the measured atomic temperature variance by more than three times, giving a more reliable result.

\end{abstract}

\maketitle


Experiments with trapped quantum degenerate gases are an important platform for the study of precise measurement~\cite{PhysRevLett.117.138501,PhysRevA.90.063606,RevModPhys.81.1051} and many-body quantum physics.\cite{RevModPhys.80.885,Schreiber842} In these experiments, the  atomic distribution is usually measured on light absorption images~\cite{apl10,apl16,smith2011absorption} after a certain time of flight (TOF). Precise determination of the atomic distribution is important, since it is from this that information such as  temperature, number of atoms, and density can be calculated, especially for quantum metrology,\cite{PhysRevLett.111.143001}  measurement of physical parameters,\cite{PhysRevA.87.053614,PhysRevA.77.012719} and investigation of phase transitions or dimensional crossover.\cite{PhysRevLett.114.230401,PhysRevLett.106.105304} In practice, there always exist some noise signals that limit the precision of measurement. Following a proposal to improve the quality of absorption images~\cite{PhysRevA.82.061606,Li07,Gunter954,NJPimaging} through the construction of an ideal light image, we present an algorithm and study and optimize its performance at the photon shot noise level, so that it can be used in the study of  cold atomic gases,\cite{PhysRevLett.107.103001} molecules,\cite{PhysRevA.81.061404} and  neutral plasmas.\cite{CESprl04}

In the conventional absorption imaging process, three pictures are taken in a cycle: first an atom image $A$ taken from the atomic cloud, then  an image $R$ taken to record the light distribution, and finally a dark-field image $G$ recording the background signal. These three pictures give the optical density $D(x,y)$ in the $xy$ plane as
\begin{equation}
D(x,y)=-\log\frac{A_{R}(x,y)-G(x,y)}{R_{R}(x,y)-G(x,y)}.
\end{equation}
The two-dimensional atomic distribution is given by $D/\sigma$, where $\sigma$ is the cross-section.
\bl{Note that the subscript $R$ on $A_{R}$ and $R_{R}$  indicates that these are  the real images that we take, rather than the values that we use in the calculation, from which the mean values $\overline{R}$ and $\overline{G}$ have been subtracted (see below).}

Besides the absorbed light due to the presence of atoms, there is still an inevitable difference between the images $A$ and $R$, which is difficult to  eliminate by adjusting the optical path. This difference  results in a fringe-type noise in the detected atomic distribution.  An effective way to reduce this fringe is to construct an ideal light distribution $A^{\prime}(x,y)$ instead of the light image $R$ for the calculation.\cite{PhysRevA.82.061606,Li07} In this approach, we construct the ideal light image based on the light distribution in an edge area by applying  principal component analysis (PCA).\cite{abdi2010principal} To minimize the noise signal in the detection, we develop an optimized fringe removal algorithm (OFRA) that reduces the noise signal  nearly to the theoretical limit of $1/\sqrt{2}$ of the photon shot noise. We carry out a qualitative study of the fringe removal algorithm, present a method to construct a light image with  optimized parameter selection,  analyze its limitations, and then compare the final result for the noise signal with the theoretical limit.

We construct the ideal light distribution of $A^\prime$ and obtain the fringe-removed atomic distribution in three steps as shown in Fig. 1. In the algorithm, the mean dark-field distribution $\overline{G}$ and the mean light field $\overline{R}$ are subtracted from both the light field and atomic absorption signals \bl{as $A=A_{R}-\overline{R}-\overline{G}$, and $R=R_{R}-\overline{R}-\overline{G}$}. Therefore, in the following, the light distributions we consider are all free of the background signals and the mean light distribution. \bl{Our first step is to decompose the images $R_i$ into a set of orthogonal basis vectors $P_j$}, with $i$ the serial number of the reference and $j$ the serial number of the basis. The second step is to reconstruct a light image $A^{\prime}$ that is similar to the atom image $A$. Finally, we obtain the atomic distribution with the given absorption image $A$ and the light image $A^{\prime}$.

\begin{figure}[htbp]
\begin{center}
\includegraphics[width=0.8\linewidth]{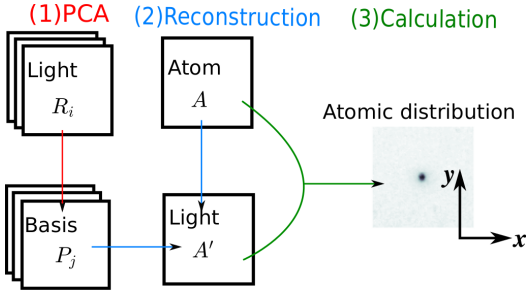}
\end{center}
\caption{Schematic of the fringe removal process:  (1) the PCA process in which the light images $R_i$ are transformed to  base images $P_j$; (2) the reconstruction process in which the atom image $A$ and the base images  $P_j$  form the ideal light image $A^{\prime}$; (3) the calculation of the atomic distribution from the atom image $A$ and the constructed light image $A^{\prime}$.}
\end{figure}

In the following, we give details of each step. In step (1), we take as references $n$ images of light $R_i$ in the experiment without changes in the experimental controls. The images are detected every 40\,s, and each  contains a different type of interference pattern arising from interference between the probe beam and its  light reflected from the optical elements. We then determine the extent to which atoms may exist based on specific physical problems. A certain region around the internal area is chosen and used to implement the algorithm. We refer to this area as an edge area, and for each $R_i$, we reshape the pixels in the edge area to \bu{a column vector $u_i$}. The dimension $l$ of $u_i$ is the number of pixels in the edge region and  is typically much larger than the number of  reference images, $n$. The edge area is chosen to exclude  information about atoms so that the final result will only reduce the fringe signal and not influence the real atomic distribution.
Following the PCA process, by defining $u_{m,h}$ as the value of the $h$th pixel of the vector $u_{m}$, a covariance matrix is given by $S_{mk}=\sum_h u_{m,h}u_{k,h}$. The eigenvectors of $S$, which can be calculated using  singular value decomposition, form a set of basis vectors $v_j$.

\bl{After obtaining the set of basis vectors $v_i$ correspond to a certain light distribution in the chosen edge area, the next step} is to construct the light distribution in the internal region. For each edge vector $u_i$, there is a corresponding full image $R_i$. \bl{If we write $v_j$ in the form $v_j=\sum_i c_{ij} u_i$, then $P_j=\sum_ic_{ij} R_i$, with the same coefficients $c_{ij}$.}

The equation is overdetermined, since the dimension $l$ of the vectors $v_j$ and $u_i$ is larger than the number of indices $i$ and $j$, so we can only solve for the least squares solutions of $c$. \bl{That is, for the final solution $c$, we have $v^{\prime}_j=\sum_i c_{ij} u_i$, and $\sum_j(v^{\prime}_j-v_j)^2$ attains its minimum value.} Therefore, in the first step, we can obtain a set of basis vectors $v_j$, as well as the corresponding full images $P_j$.

In step (2), our aim is to construct an ideal light image $A^\prime$ from the basis vectors $P_j$ based on the absorption image $A$. With the defined edge area, the atom image $A$ also gives an edge vector $u_0$. Then its expansion coefficients are given by the scalar products of the vector $u_0$ with the vectors $v_j$, \bu{$\omega_{0j}=u_0^{\rm T} v_j$ (where T denotes the transpose)}, and are used are used to reconstruct the ideal light image as
 $A^\prime=\sum_j \omega_{0j}P_j$.
\bl{
Finally, in step (3), by rewritting Eq.~(1) as
\begin{equation}
D(x,y)=-\log\frac{A_{R}(x,y)}{A^{\prime}_{R}(x,y)},
\end{equation}}
we can obtain an  atomic distribution that is now fringe-free.

After this brief introduction to the OFRA, we now demonstrate how it can be optimized. In step (1), we first need to choose an optimized edge area where the calculation is performed. In step (2), \bl{we  select the basis vectors as the principal components}.


To find the best way to select the edge area, we test the performance of our algorithm with different edge regions. The algorithm's performance is measured by the atomic distribution derived from a pure light field image outside the reference set. Because there is no absorption, the result we obtain from the algorithm is completely a noise signal. Thus, the mean value of the noise signal in the region surrounded by the chosen edge area can be considered as the evaluation criterion for the algorithm.

\begin{figure}[ht]
\centering
\includegraphics[width=\linewidth]{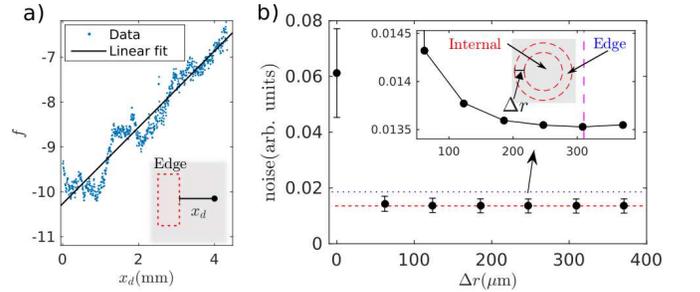}
\caption{\bl{(a) The blue dots show the decay rate of the noise signal $f$ with  distance as defined in the  text, and a linear fitting (black solid line) is also presented. The inset shows the definition of the distance $x_d$. (b) Variation of the performance of the fringe removal algorithm with radius $\Delta r$ of the  edge region chosen for calculation. The inset shows an amplified view of the noise signal together with the definition of $\Delta r$.}}
\end{figure}

Reconstruction of the entire light field image  depends on the calculation in the edge region. Therefore, the validity of the algorithm is based on the assumption that the similarity between the reconstructed image and the ideal light image in the edge region can be extended to the central region. To better understand this algorithm, we measure the spatial relevance of the probe light field distribution, by taking as the edge region a rectangle with $100\times 600$ pixels (the side length of each pixel is 6.8\,$\mu$m in our case) and calculating the dependence of the average noise signal on distance as $F_1(x_d)$ in the same range of $y$, with $x_d$ the distance from one point to the edge region in the $\hat{x}$ direction as shown in Fig.~2(a).

For comparison with the noise signal, we also calculate the mean noise signal between two arbitrary light images in our system, denoted by $F_0(x_d)$. In practice, we take the mean of 100 individual results from different light images that are not used as references, and a ratio $f$ is defined as follows to represent the decay rate of the noise signal:
\begin{equation}
f=10\log_{10}(F_1/F_0).
\end{equation}
Here, the influence of the light intensity distribution and its variations in amplitude in different images are excluded, and the results are presented in Fig.~2(a).

For pixels near the chosen edge region, the noise signal can be reduced by one order of magnitude on average. When the distance $x_d$ grows, the magnitude of the noise starts to increase. We fit the noise attenuation rate $f$ linearly, and it drops to $1/e$ of the best value at a distance $0.51\pm0.01$\,mm. This is the \bl{reconstructed distance} for the light field of the imaging laser, which is related to the spatial distribution of different types of fringes in the system. In addition, we also note that there are points with extremely high noise (the peak near $x_d=2$\,mm). \bl{These are due to the presence of dust or small defects on the mirror and  can be avoided by adjusting the experimental parameters or the imaging light path slightly to find a position where these small spots or defects do not exist in the region where atoms would be detected.}

After measuring the \bl{reconstructed distance}, the next question is how the edge area  influences the performance of the fringe removal algorithm.
Considering the atomic distribution in real space, an annular region is chosen as the edge area to construct the basis. The inner diameter of this region is chosen as 432.6\,$\mu$m (70 pixels) and the outer diameter as $432.6\,\mu$m + $\Delta r$. $\Delta r$ is changed from 10  to 70 pixels.

In Fig. 2(b), the conventional result calculated without the algorithm is given with coordinate $\Delta r=0$. When the algorithm is used, the noise signal drops greatly, by more than four times, from 0.06 to about 0.014. \bl{With  increasing  $\Delta r$}, the performance of the algorithm initially improves and becomes best  at about $\Delta r=46\times 6.8\,\mu$m. This  is close to half the distance between fringes in the most prominent component at 310\,$\mu$m, as shown in Fig.~2(b) by the magenta dashed line. Here, we have verified the general conclusion that the light image $A^{\prime}$ can be well reconstructed when \bl{at least the length of one cycle of the most significant fringe shown in Fig. 3 is included in the edge area as given in Fig. 2(b). Considering the edge surrounding this area from both sides, we need a $\Delta r$ that is about half the length of one cycle of the fringe.}

After selecting the edge region, in step (2) another factor that influences the performance of the algorithm is the way in which the number of prominent components is chosen. The application of PCA has several advantages, one of which is that PCA  gives  components arranged in order of importance, which can be utilized to remove redundant information. There is also a method known as 2DPCA~\cite{2dpca} that has been  suggested for problems like face recognition. However, in our case, the atom is distributed in a central area, and the application of  2DPCA cannot speed up the algorithm further compared with  ordinary PCA applied only in an edge area. Therefore, we implement  PCA with all the pixels in the edge region transformed into a one-dimensional vector. In Fig.~3, the eigenvalue of each component is presented on a logarithmic scale, together with some of the corresponding pictures of $P_j$. The first principal component $P_1$ shows a clear fringe appearance and is the source of the greatest noise in our imaging system. \bl{Then, for $P_{100}$, the fringe spacing becomes smaller, and the fringe distribution can be seen clearly only in the upper left corner of the figure. The last component $P_{310}$  shows only random noise.} The performance of the algorithm is also measured by the noise signal calculated from the light image without any atomic distribution. By preserving different numbers of base vectors, the optimal performance in our system was achieved by keeping the basis vectors with $j\leq 150$. This is also consistent with the visualization of the basis image: the value $j=150$ is approximately the threshold at which $P_j$ changes from regular fringe to random noise. \bl{It should be noted that  the method for selecting the optimal principal element discussed here was also applied to obtain the results  of step (1) shown in Fig.~2(b). Thus, it provides the best performance of the OFRA.}

\begin{figure}[htbp]
\begin{center}
\includegraphics[width=\linewidth]{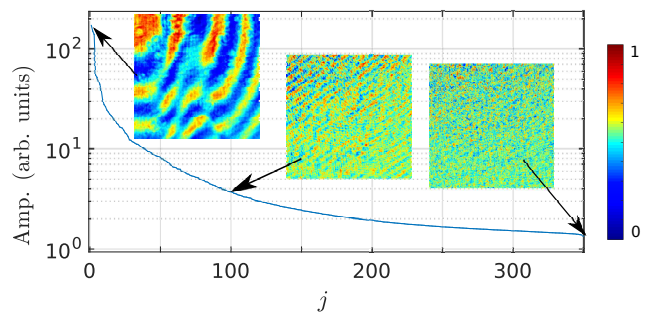}
\end{center}
\caption{The amplitude of the eigenvalue of $v_j$  calculated as the basis in PCA. The three pictures from left to right show the corresponding full images of $P_j$ with $j=1$, 100, and 310, respectively.}
\end{figure}

\bl{We now compare the noise signal in our final result with the theoretical limit. In our system, the CCD typically records about 1000 photons for each pixel, and the photon shot noise is approximately 30 photon counts. This noise exists in both light images and atom images for conventional methods, and the final noise is multiplied by a factor of $\sqrt{2}$. Using Eq.~(1), we obtain a noise signal of 0.0185, as shown in Fig. 2(b) by the blue dotted line. In comparison, our OFRA gives a result of 0.0135, reducing the noise by a factor of $1/\sqrt{2}$, as shown by the red dashed line, because the reconstructed image is a weighted average over a set of light images and we have also dropped the noisy components. Thus, the photon shot noise is strongly suppressed in the reconstructed light images.\cite{PhysRevA.82.061606,NJPimaging}}

\begin{figure}[htbp]
\centering
\includegraphics[width=0.8\linewidth]{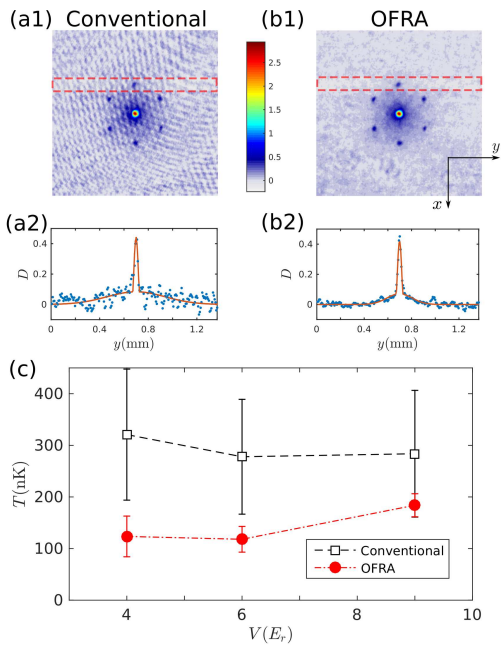}
\caption{Comparison of the TOF images obtained from a conventional calculation  (a1) and using the fringe removal algorithm (b1). (a2) and (b2) show the one-dimensional atomic distributions obtained by summing  the distribution in the rectangular area of the TOF images over the horizontal direction. Dots are experimental measurements and the red solid line gives the bimodal fitting curve. (c) Temperature fitted from the scattering peaks as shown in the second row at different lattice depths $V$ with (open rectangles) and without (red circles) the OFRA.}
\end{figure}

To show the effectiveness of the algorithm, we also apply it to measurement of physical parameters in the complex case. We begin with a condensate in a hybrid trap, as in our previous work.\cite{Hu2018Ramsey,PhysRevA.94.063603} During the detection process, the density distribution of the atom cloud is measured after a 31\,ms TOF from an absorption image. The nearly homogeneous probe laser is typically tuned to resonance with the atomic transition and the laser pulse lasts for about 50\,$\mu$s. The resulting light field signal is imaged onto a CCD camera. The exposure time can be controlled with a V++ control program by adding a TTL trigger signal. In practice, we choose as an example a triangular lattice in which the atoms are distributed over an area much larger than that of a released condensate. For each of the six scattering peaks, as shown in Fig.~4, the atomic population is small, making  detection more difficult, as shown in Figs.~4(a1) and 4(b1). Fortunately, the radius of the atomic distribution in the TOF image is within 0.432\,mm, which is  smaller than the calculated distance of 0.51\,mm at which we can still reconstruct the light image with  good performance.

The experimental configuration is described in our previous work.\cite{zhou2018shortcut} In the experiment, we raise the depth of the lattice adiabatically to a final value $V$ and keep the atoms in the lattice potential for 20\,ms before the measurement. For the triangular lattice, there are several parameters characterizing the system,\cite{BCnjp10} including visibility, condensate fraction, and temperature. We perform a bimodal fitting to the scattering peaks in the direction perpendicular to the center as shown in Fig.~4(a) for the conventional  calculation and in Fig.~4(b) for the OFRA by summing  the atomic distribution within the red box. The bimodal curve consists of a \bu{Boson enhanced Gaussian} distribution representing the thermal component and an inverse parabolic curve representing the condensed component. For these two components, we measure the column densities along the imaging axis, which are given by
\begin{equation}
\begin{aligned}
n_\mathrm{th}(x) &= \frac{n_\mathrm{th}(0)}{g_2(1)}\,g_2\!\left(\exp\!\left[-\frac{(x-x_0)^2}{\sigma_T^2}\right]\right),\\[6pt]
n_{c}(x) &= n_{c}(0)\max\!\left[1-\frac{(x-x_0)^2}{\chi^2} \color{blue}, 0\right]\color{blue}^{3/2}.
\end{aligned}
\end{equation}
There are five fitting parameters: the amplitudes $n_\mathrm{th}(0)$ and $n_{c}(0)$ of the two components, the widths $\sigma_T$ and $\chi$ of the two components, and the center position $x_0$ of the atomic cloud. The Bose function $g$ is defined as \bu{$g_j(z) = \sum_{i=1}^\infty z^i/i^j$. And the `max' denotes that when the value before the comma is lower than zero we should take the value as zero.} In practice, we apply least squares fitting of $n_\mathrm{th}(x)+n_{c}(x)$ to the real distribution that we measured. From the fit, we can obtain the numbers of atoms and the widths of the two components separately. Measurements are performed at different lattice depths and for each lattice depth, with 30 experiments for each case. 

After obtaining the width $\sigma_T$ of the thermal component, the temperature is given by
\begin{equation}
T=\frac{1}{2}\frac{M\sigma_T^2}{t_\mathrm{TOF}^2k_B},
\end{equation}
with $M$ the atomic mass and $t_\mathrm{TOF}$ the TOF time.\cite{PhysRevA.55.R3987}

For parameters such as the number of condensed atoms, the measurement is less affected by the fringe. However, for the determination of temperature, which is proportional to the width of the \bu{Boson enhanced Gaussian} distribution, the influence of the fringe on the fitting is much more prominent. Figure~4 shows the temperature obtained from the TOF images with and without the OFRA. The TOF images are shown in Figs.~4(a1) and 4(b1), while the integrated one-dimensional atomic distributions are shown in Figs.~4(a2) and 4(b2) for the conventional calculation and for OFRA, respectively. In Fig.~4(c), the open rectangles show the temperature obtained from the conventional calculation. This temperature is in the region of 300\,nK, which is far higher than the initial temperature of a Bose--Einstein condensate (about 90\,nK). \bl{During the loading process, the lattice potential is turned on adiabatically over a period of 80\,ms, and this will lead to a limited heating effect. Even if we assume that the temperature of our system can be increased by a factor of three, the proportion of condensed atoms should be reduced significantly, so the result from the conventional calculation is still not consistent with observations. When the OFRA is applied, as shown by the red circles, the temperature is obtained with a much smaller variance and has a much more reasonable value.} For a lattice depth  $V=4E_r$, the  temperature obtained is 123.5\,nK, and for $V=9E_r$, it grows to 183.9\,nK. Comparison of these two results shows that in the case of a small  number of atoms, when fitting physical quantities such as the temperature, we can obtain a reliable result only if the fringe removal algorithm is used.

Compared with other fringe removal methods,\cite{PhysRevA.82.061606,NJPimaging,Li07}  OFRA is optimized and faster in two aspects. First, our study shows that the area used for the calculation has an impact on the final result of the algorithm. By studying the effect of reconstructing the light field at different distances, we provide a criterion that can be extended to other systems, and the optimal calculation scheme needs to include only one cycle of the fringe pattern of the first principal component in space. Second, by implementing  PCA, we keep only the principal components containing information on the interference fringe. Both of these improvements could greatly reduce the calculational load while still achieving the best performance. It is also noteworthy that in the absence of our proposed optimization condition, the introduction of a larger edge area or more non-prominent bases will make the algorithm worse in terms of both effect and speed. This is known as the over-fitting effect, and it is more significant when the number of reference images is small. Thus, our study can be applied to precision measurement and other fields to more effectively obtain  information on atomic distributions.

In summary, we propose a method for generating an ideal light field in the absorption imaging process to avoid noise due to changes in the light field. We optimize the parameters in the algorithm based on the amplitude of the noise signal to achieve the best performance. The spatial reconstruction distance of the probe light field is also given, which is not only a prerequisite of the algorithm, but also provides the basis for  selection of the calculation region. As shown in the case of a triangular lattice, with this algorithm, we can measure parameters that cannot be obtained with high precision by conventional methods. The OFRA that we have developed is also easy to implement in an  absorption image-based experiment, requiring only some algorithmic modifications without any changes to the experimental system.

We thank Lan Yin and J\"org Schmiedmayer for helpful discussions.
This work is supported by the National Key Research and Development Program of China (Grant No. 2016YFA0301501) and the National Natural Science Foundation of China (Grant Nos. 11334001, 61475007, 61727819, and 91736208). W. M. Liu acknowledges support from the National Key R$\&$D Program of China under Grant No. 2016YFA0301500, the NSFC under Grant Nos. 11434015, 61227902, \bu{61835013,} and the SPRPCAS under Grants Nos. XDB01020300 and XDB21030300.

\bibliography{citation}
\end{document}